\documentclass[journal]{IEEEtran}
\usepackage{amsmath,amsfonts,amssymb}
\usepackage{array}
\usepackage[caption=false,font=normalsize,labelfont=sf,textfont=sf]{subfig}
\usepackage{textcomp}
\usepackage{stfloats}
\usepackage{subcaption}
\usepackage{url}
\usepackage{verbatim}
\usepackage{graphicx}
\usepackage{cite}
\usepackage{color}

\usepackage{xspace}
\newcommand{\umux}{$\mu$mux\xspace} 

\begin{document}

\bstctlcite{IEEEexample:BSTcontrol} 

\title{Development of microwave multiplexed readout for athermal phonon TES-based detectors}

\author{Tran Tsan, Douglas A. Bennett, Daniel T. Becker, Rebecca Carney, Johnathon D. Gard, John A.B. Mates, Joel N. Ullom, Michael R. Williams, John C. Groh
\thanks{Tran Tsan, John C. Groh, and Michael R. Williams are with the Physics Division, Lawrence Berkeley National Laboratory, Berkeley, CA 94720, USA.}
\thanks{Douglas A. Bennett and John A.B. Mates are with National Institute of Standards and Technology, Boulder, CO 80305, USA.}
\thanks{Daniel T. Becker and Jonathan D. Gard are with the Department of Physics, 
University of Colorado, Boulder, CO 80309, USA.}
\thanks{Rebecca Carney is with the Engineering Division, Lawrence Berkeley National Laboratory, Berkeley, CA 94720, USA.}
\thanks{Joel N. Ullom is with the National Institute of Standards and Technology, Boulder, CO 80305, USA and the Department of Physics, University of Colorado, Boulder, CO 80309, USA.}
}

\maketitle

\begin{abstract}
Continued advancements in sub-eV calorimetric measurements are being enabled through the athermal phonon detection channel, with applications spanning a diverse array of fields, including low-mass dark matter searches, coherent elastic neutrino-nucleus scattering, and quantum information science. While Transition Edge Sensors (TESs) have achieved the lowest thresholds to date, they face significant scalability bottlenecks due to the inherent constraints of conventional SQUID readout architectures. To address this limitation, we present an update on R\&D to develop practical, highly multiplexed, microwave readout for athermal phonon TESs which can enable a dramatic increase in both sensor count and target mass, with a focus on sub-GeV dark matter searches as a target application.  This work leverages advances from highly scaled mm-wave, x-ray, and gamma-ray TES systems, adapting them to the unique needs of low-threshold rare event measurements.  We quantify the basic readout functionality and noise level in our dilution refrigerator testbed and demonstrate operation of a variety of TES devices with saturation powers significantly lower than mm-wave and x-ray devices.  We demonstrate proof-of-concept multiplexed readout of background event pulses, bound the readout's contribution to the parasitic TES background load to $<$1~fW, and discuss a cryogenic inline RF filtering scheme.  These results demonstrate practical solutions to several of the key challenges that must be solved before a next-generation athermal phonon detector at scale is realized.

\end{abstract}

\begin{IEEEkeywords}
athermal phonon sensing, TES, SQUID, multiplexed readout, microwave resonator
\end{IEEEkeywords}

\section{Introduction}
\label{sec:intro}

\IEEEPARstart{L}{ow}-threshold athermal phonon detectors are being developed for a range of applications requiring sensitivity to extremely small energy depositions, including low-mass dark matter searches \cite{Alkhatib_2021, Bui_2025}, coherent elastic neutrino-nucleus scattering \cite{AUGIER2023}, and quantum sensing \cite{Fink2024}. In keV--GeV dark matter searches, where constraints on interaction cross sections are orders of magnitude weaker than at WIMP-scale masses, probing lower particle masses requires sensitivity to energy depositions at the eV scale and below. This motivates the development of highly sensitive superconducting sensors operating at cryogenic temperatures. Transition-Edge Sensors (TESs) coupled to athermal phonon absorbers have demonstrated exceptional energy resolution: recent work  has demonstrated a phonon energy resolution of 0.361 eV at 3.06 eV \cite{Bui_2025}. While these detectors provide excellent sensitivity, scaling them to larger target masses and channel counts presents a significant readout exercise. Current implementations use $\mathcal{O}(10)$ individual DC SQUID readout channels, for which the cryogenic wiring and readout complexity become increasingly difficult as detector counts grow. Future experiments envision scaling to $\mathcal{O}(10^3)$ sensors or more, motivating the development of a scalable multiplexed readout architecture.

Multiplexed TES readout has been extensively developed for applications including mm-wave astronomy \cite{McCarrick_2021, Groh_2025}, x-ray spectroscopy \cite{Szypryt_2023}, and $\gamma$-ray spectroscopy \cite{schreiber2025}. In particular, microwave SQUID multiplexing ($\mu$mux) combines GHz-scale readout bandwidth with high multiplexing factors, with recent system demonstrating multiplexing factors approaching 2000 TES channels per readout unit \cite{Groh_2025}. In a $\mu$mux system, each TES is inductively coupled through an rf-SQUID to a unique GHz-frequency microwave resonator. The resonators share a common transmission line, and flux ramp modulation linearizes the SQUID response. 

However, TESs optimized for low-energy phonons differ significantly from those in existing \umux deployments, introducing new constraints. For example, athermal phonon TESs can be physically larger and distributed over a larger volume, imposing different constraints on the interconnects between sensors and readout circuitry. They also generally operate near 10 mK, substantially below the typical operating temperature ($\sim$100 mK) of many existing $\mu$mux systems. Furthermore, athermal phonon TESs optimized for dark matter searches often have a saturation power, $P_{sat}$, of $\mathcal{O}(1-10)$~fW, several orders of magnitude less than TESs optimized for CMB, x-ray, and $\gamma$-ray detection. Consequently, parasitic power negligible in conventional \umux systems may therefore contribute appreciable background load or even saturate these detectors. Understanding and controlling this background is an important part of adapting $\mu$mux to athermal phonon TES detectors. 

We present progress toward developing $\mu$mux readout for athermal phonon TESs for low-mass dark matter detection. We envision that these results could enable future scaled-up experiments using sensors such as those used by the TESSERACT collaboration \cite{Bui_2025}.  We describe experimental design modifications (Section~\ref{sec:experimental_preparation}), evaluate our readout performance at 10~mK (Section~\ref{sec:readout_checks}), demonstrate proof-of-concept TES readout (Section~\ref{sec:TES_operation}), and discuss challenges and conclusions (Section~\ref{sec:discussion}).

\section{Experimental Design and Setup}
\label{sec:experimental_preparation}

In adapting \umux readout for dark-matter athermal phonon TESs, we aimed to minimize modifications to the existing architecture and use established components wherever possible. The readout uses a \umux chip originally developed for the Simons Observatory (SO) \cite{dober2021microwave, McCarrick_2021}.  A software-defined radio readout system based on an RF system-on-a-chip (RFSoC) provides microwave tone generation and flux-ramp demodulation, mixed up to the 4--8 GHz band \cite{Gard2018}, with an externally-supplied local oscillator.\footnote{Keysight PNA-X}  The flux ramp waveform is supplied by an arbitrary waveform generator.\footnote{Keysight 33509B}  For representative TESs, we used a collection of tungsten ``bare" TESs and quasiparticle-trap-assisted electrothermal-feedback TESs (QETs), with transition temperatures of approximately 48~mK and anticipated saturation powers spanning $\sim$10-800~fW~\cite{watkins2023}. This range allows us to test the readout under different power-loading requirements and provides a diagnostic of parasitic background loading. For this proof-of-concept, we wired up six detector channels, including two QET channels on the same detector chip, enabling simultaneous multiplexed readout of correlated phonon pulses. 

\begin{figure}[htbp]
\begin{center}
\includegraphics[width=0.95\linewidth]{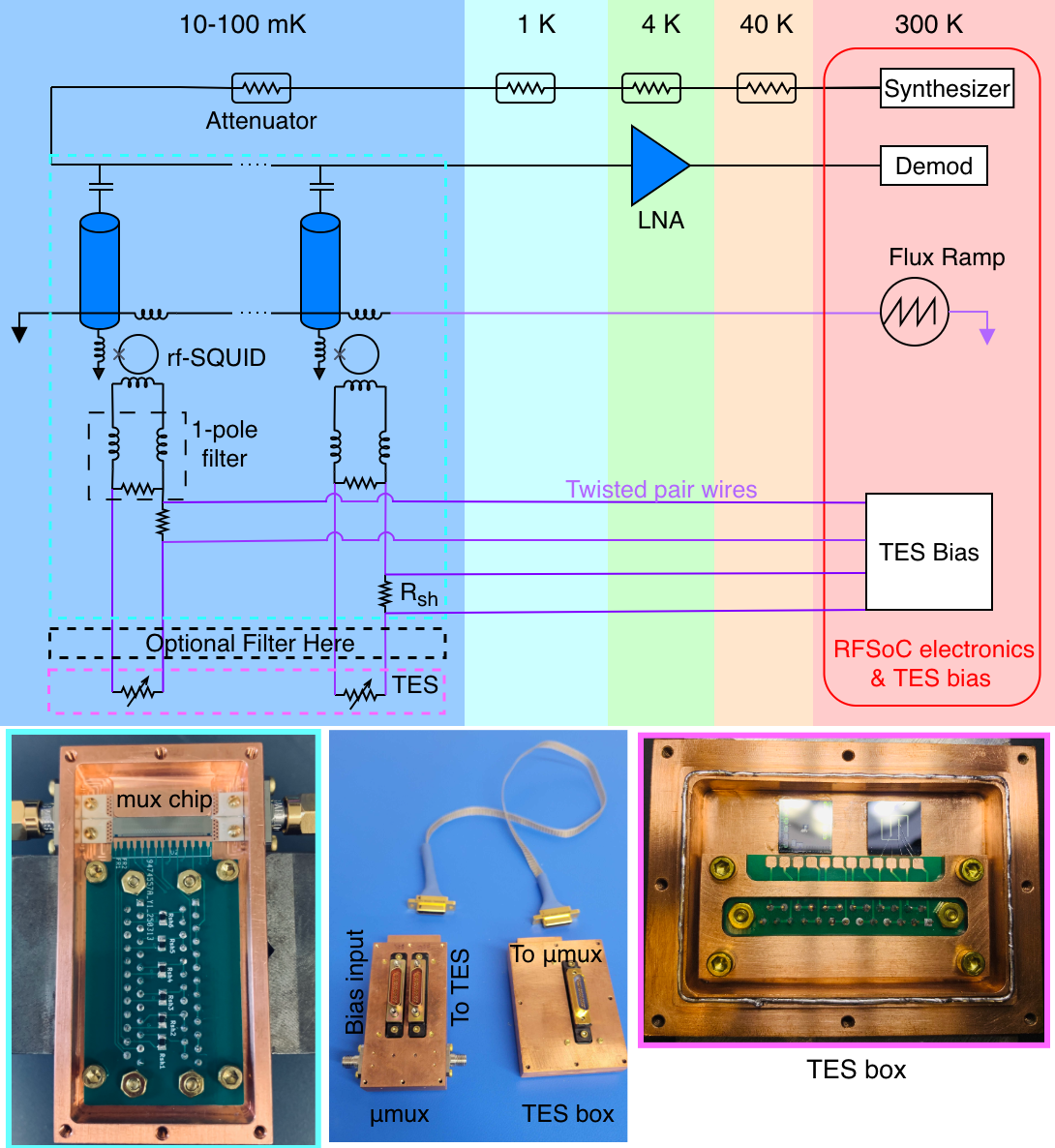}
\end{center}
\caption{Schematic (top) and photographs (bottom) of our R\&D setup. The basic \umux architecture follows \cite{McCarrick_2021, dober2021microwave}, with modifications for athermal phonon TES testing. The readout and detectors are housed in separate enclosures connected by twisted pair wires and 25-pin micro-D connectors. This separation permits optional inline filtering to reduce readout leakage into the detectors.}	
\label{fig:setup}
\end{figure}

To accommodate the larger physical footprint of the detector chips, the TESs and \umux are housed in separate enclosures connected by twisted-pair wiring and micro-D connectors, with two pins used for each TES lead to reduce parasitic resistance. The relatively higher TES resistance, $\mathcal{O}(0.1) \Omega$, relaxes the parasitic constraints, allowing the sensors to be separated from the readout using this extended wiring schematic. The twisted-pair connections serves as a proxy for any extended wiring that may be required between the readout and sensors in a scaled up experiment. Separating the two assemblies also provides a convenient location for possible inline filtering between the sensors and readout. Each TES has an independent bias line, allowing devices with different saturation powers to be characterized individually. We note however that it is routine to chain the bias resistors in series for large TES arrays to reduce the number of bias wires \cite{mates2011}, consistent with the multiplexing paradigm. Figure \ref{fig:setup} shows the two assemblies and their electrical connections. The TES box contains two 1~cm~x~1~cm chips, mounted directly to a copper surface with rubber cement. This initial mounting scheme was chosen for simplicity but is a known source of low-energy excess (LEE) events and will be improved in future iterations. We seal the box with indium wires along the enclosure interfaces to reduce infrared radiation reaching the sensors.

The \umux enclosure contains a multiplexer chip and $R_{sh}$~=~5~m$\Omega$ shunt resistors for the TES voltage biases. Two multiplexer chips were used during this study: one covering $\sim4.8-5$ GHz for the readout characterization in Section \ref{sec:readout_checks}, and a second covering $\sim4.5-4.7$ GHz for the TES measurements in Section \ref{sec:TES_operation}. Of the 65 available readout channels, six are connected to TESs in the present configuration. Both the TES and \umux boxes are mounted to the mixing chamber stage of a dilution refrigerator and cooled to approximately 10~mK. The modular two-box design also allows the \umux box to be moved to a higher temperature stage if required, while accommodating the mux chip's need for a grounded, electrically small enclosure and the TESs' need for a low-loading environment. The cryostat's RF input chain provides about 40 dB of attenuation before \umux, while the output chain provides approximately 40 dB of amplification. During nominal operation, approximately $-75$ dBm of probe tone power is delivered to the resonators, consistent with optimum power for readout noise \cite{dober2021microwave, schuster2023}. 

As discussed in Section \ref{sec:intro}, a notable concern in adapting \umux to low-$P_{sat}$ TESs is the parasitic microwave loading. The microwave probe tones used to interrogate the resonators inevitably couple through the rf-SQUID into the TESs. The \umux chip incorporates a single-pole lumped-element filter with a nominal cutoff frequency of $\sim$10~MHz between the SQUID and TES to suppress this leakage. However, the substantially lower saturation powers of dark-matter-optimized athermal phonon TESs motivate additional characterization of this leakage and, if necessary, further filtering between the multiplexer and detectors.  We identified several commercial lumped-element low-temperature co-fired ceramic (LTCC) filters\footnote{MiniCircuits LFCN-1800D+, LFCN-2850D+, and LFCN-3000D+} as candidates for additional microwave suppression. These compact, surface-mount components are smaller than the TES absorbers. Figure 2 shows the transmission of the LFCN-1800D+, which provides $>$20~dB of rejection at 10 mK across the 4-8 GHz band (with $>$30~dB achievable using the better-matched LFCN-2850D+). Additionally, the measured 4~m$\Omega$ series resistance of the 3000D+ is sufficiently small compared to the $\mathcal{O}(1) \Omega$ TES resistance to maintain stable voltage bias.  These measurements establish LTCC filters as viable options for inline microwave filtering if additional suppression is required.

\begin{figure}[htbp]
\centering
\includegraphics[width=\linewidth]{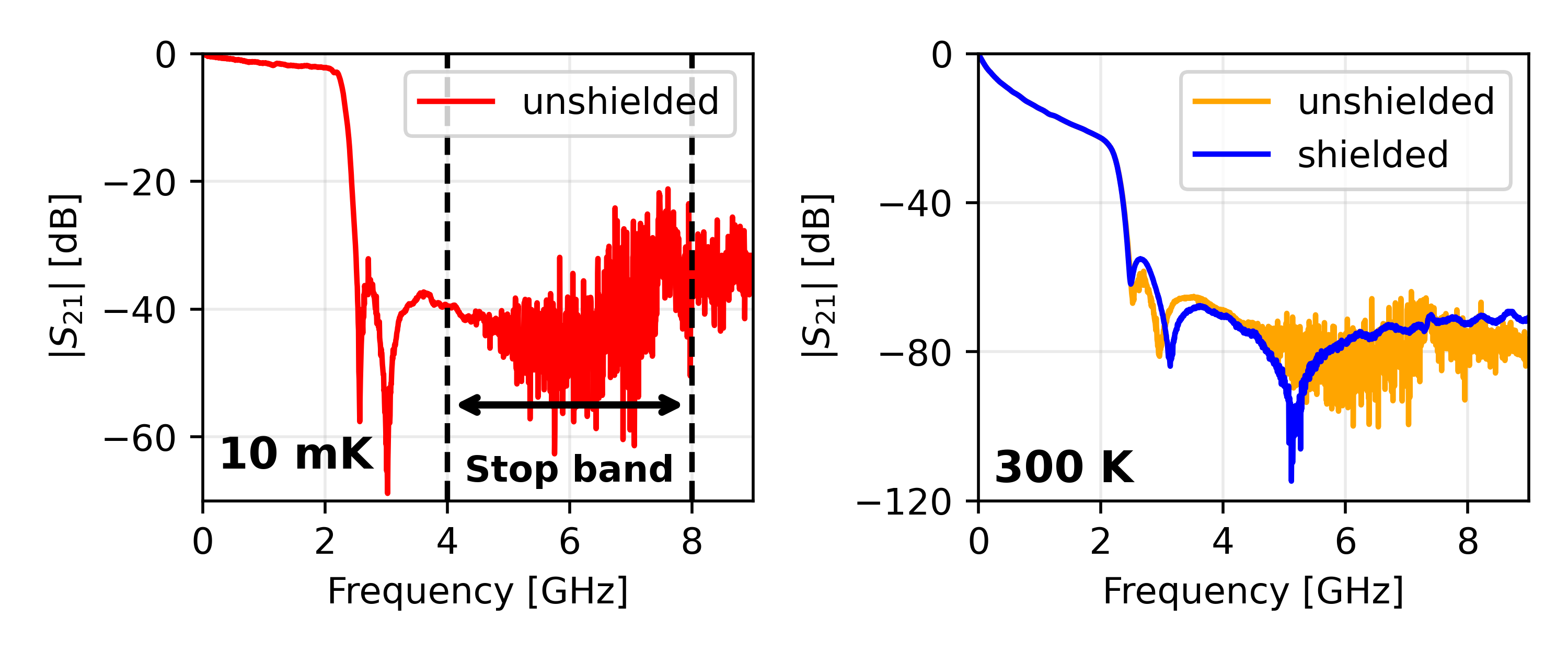}
\caption{LTCC candidate filter performance.  Left panel: cryogenic transmission when tested without any radiation shielding.  The stop band rejection is $\gtrsim$30 dB, but with oscillations due to radiative coupling to the fridge's mixing chamber can.  Right panel: room temperature verification that the high-frequency oscillations can be removed with a simple radiation shield surrounding the filter component.}
\label{fig:filter}
\end{figure}

\section{Readout System Characterization}
\label{sec:readout_checks}

\begin{figure}[htbp]
\begin{center}
\includegraphics[width=0.49\linewidth]{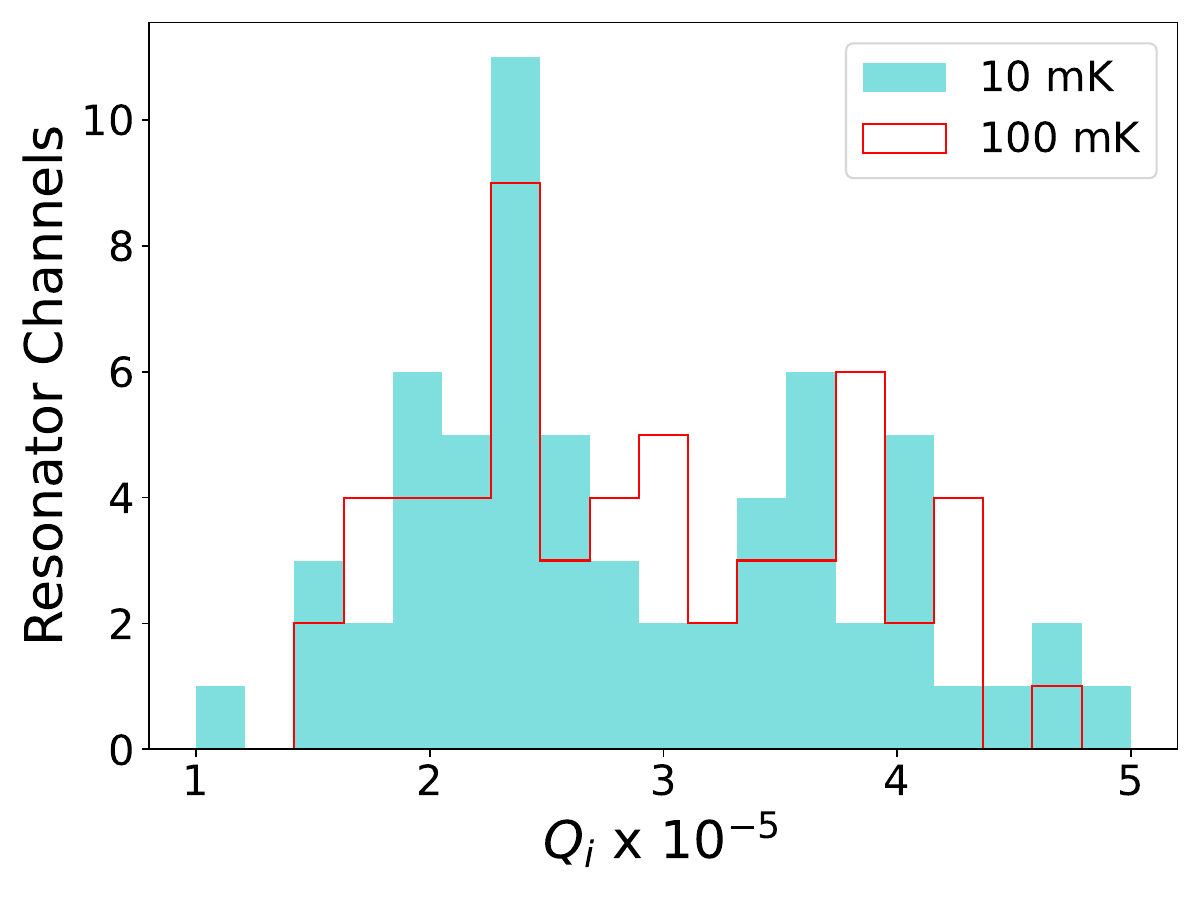}
\includegraphics[width=0.49\linewidth]{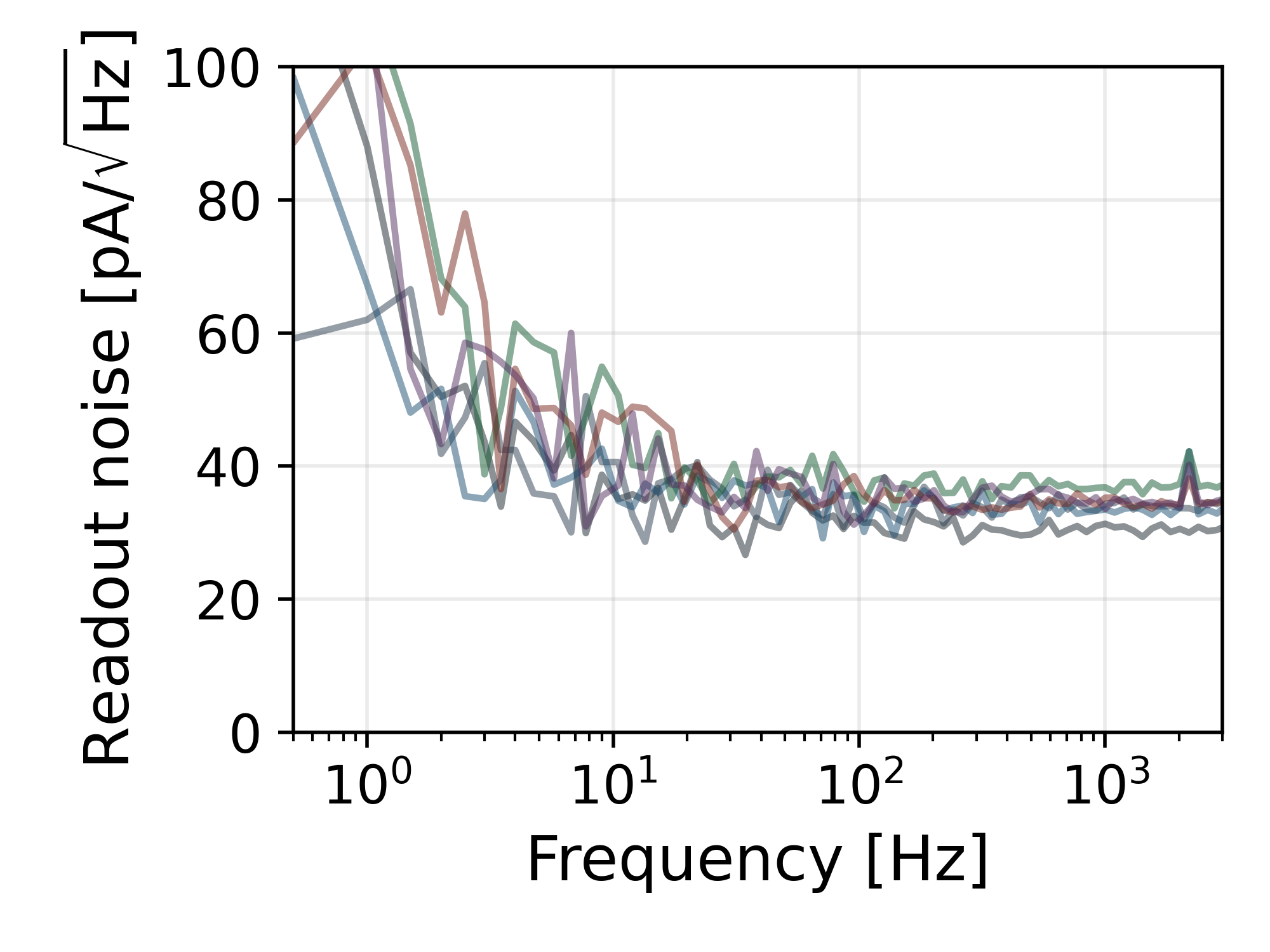}   
\end{center}
\caption{Left: distribution of the resonator internal quality factors at 10 and 100 mK. The two distributions  show that $Q_i$ does not change significantly at lower temperature. Right: power spectral density of the input current for different readout channels. The noise floor corresponds to $\approx$3.8~$\mu\Phi_0 \mathrm{/\sqrt{Hz}}$, similar to that of previously deployed systems in mm-wave and x-ray experiments.}	
\label{fig:readout_chara}
\end{figure}

Before connecting the TESs, we first verified the performance of the \umux system in our dilution refrigerator testbed.  We first characterized the multiplexer chip's resonator and SQUID parameters with a vector network analyzer and static SQUID flux bias. We focused on the resonator internal quality factor, $Q_i$, which measures the dissipation in the microwave resonators and is known to depend on the temperature. For this chip design, $Q_i$ should be above $10^5$ to avoid a noise penalty \cite{jones2024qualification}.  Figure \ref{fig:readout_chara} (left) compares $Q_i$ distributions extracted from fits to the complex readout line transmission, $S_{21}$, at 100 and 10 mK with approximately $-75$ dBm of probe tone power at the mux chip. We observed no significant temperature dependence between the operating temperature in the Simons Observatory and in target dark matter applications.

We next operated the multiplexer using the software-defined radio readout system and measured the input-current-referred noise with the TESs disconnected.  We use a $2\Phi_0$ amplitude flux ramp sawtooth waveform, discarding one SQUID oscillation per period to avoid the reset transient. The noise level is a function of the flux ramp modulation rate, and depends on the details of the local oscillator source, RF mixers, and resonator bandwidth \cite{yu2022bandwidth}.  Figure \ref{fig:readout_chara} (right) shows the noise spectra at a flux ramp reset rate of 7.1~kHz, at which the noise is minimized in our setup. The median white noise level is $34 \mathrm{ \ pA/\sqrt{Hz}}$, comparable to previously reported readout noise \cite{McCarrick_2021, Szypryt_2023}. Together, these measurements demonstrate that the readout system as implemented in our R\&D setup maintains expected performance.

\section{TES operation}
\label{sec:TES_operation}
\par
\begin{figure}[htbp]
\centering
\includegraphics[width=0.49\linewidth]{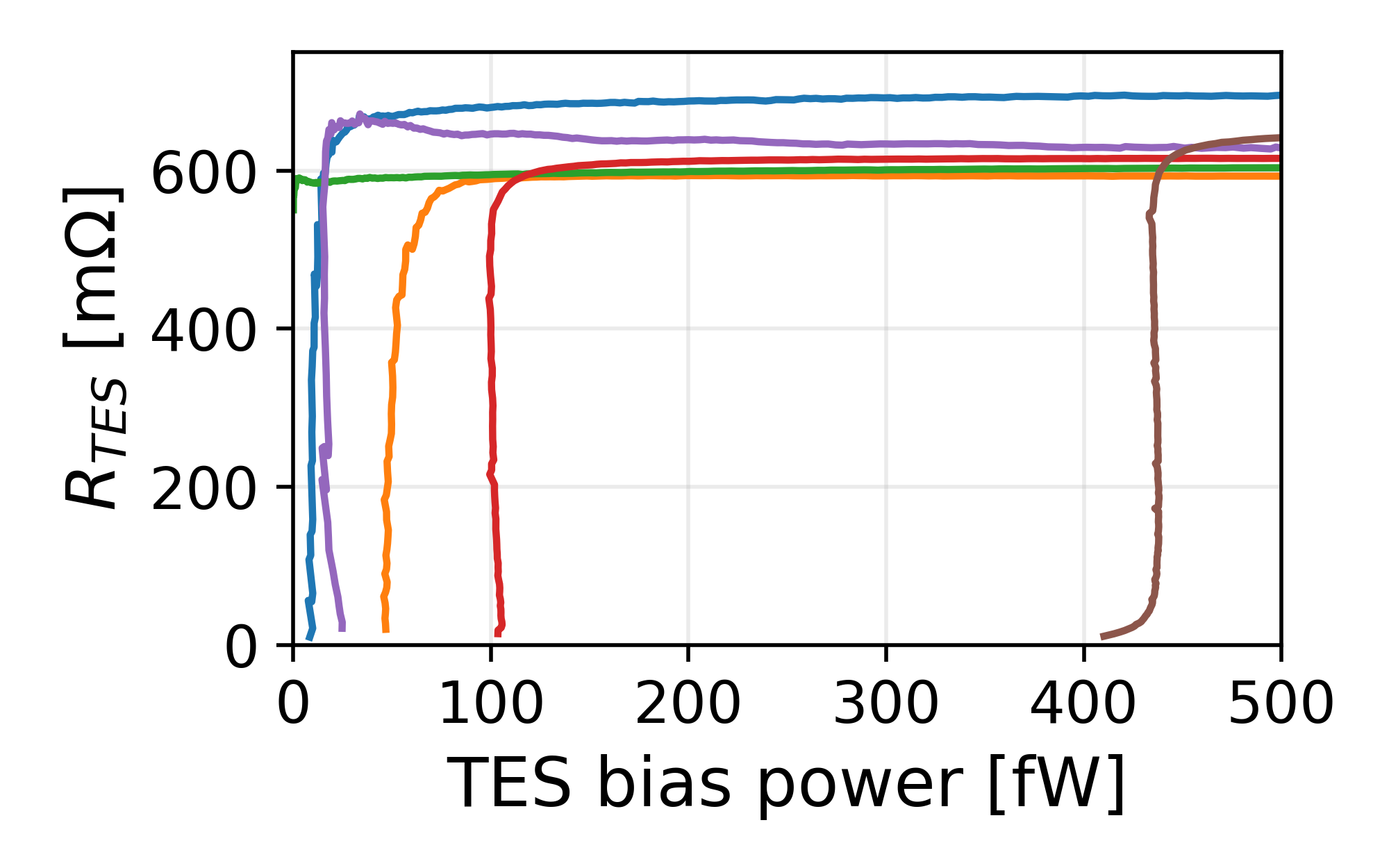}   
\includegraphics[width=0.49\linewidth]{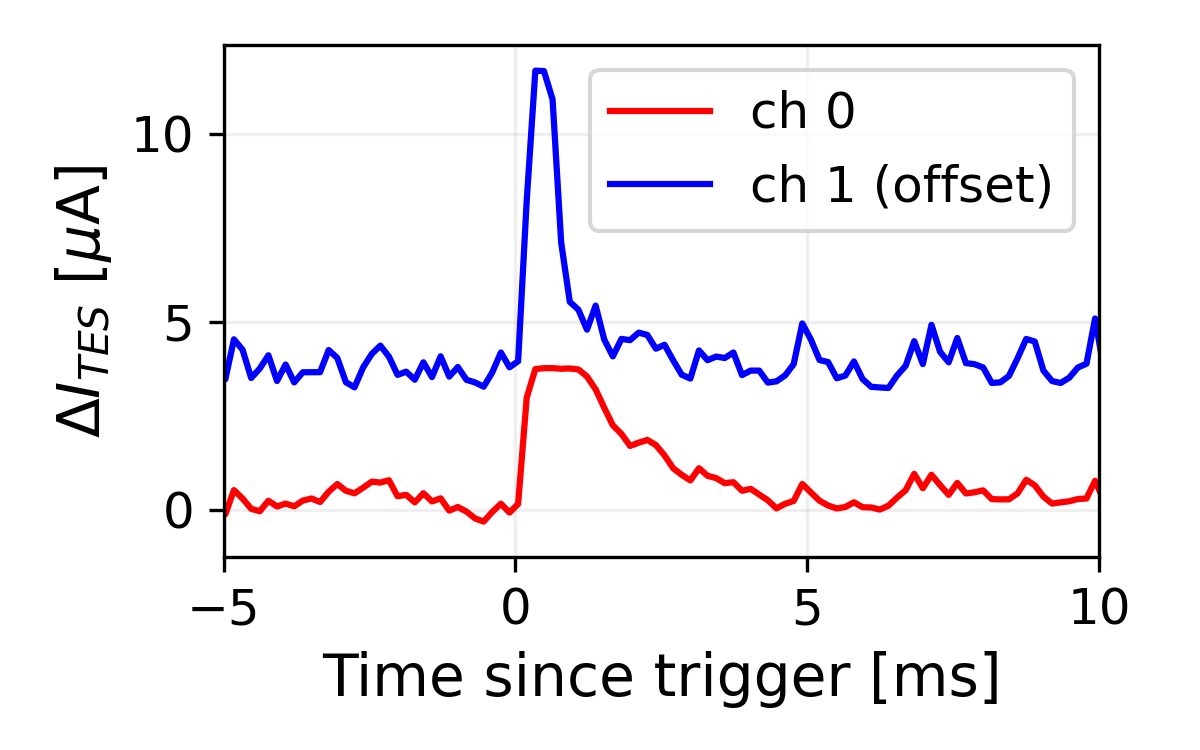}
\caption{\textit{Left:} resistance vs. bias power relationships for the six TES devices tested, demonstrating that a wide range of saturation powers are feasible with our setup.  The data points inside the superconducting transition regions are slightly contaminated by the influence of residual background pulses; the jagged shapes are therefore not expected to reflect the true zero-frequency TES behavior.  \textit{Right:} examples of correlated pulses (indicative of athermal phonon bursts) detected with two QET channels on the same chip tuned into their superconducting transition.}
\label{fig:IV_pulses}
\end{figure}

 
The electrical properties of the TES devices in our setup were characterized at DC through a bias voltage sweep.  As shown in the left panel of Figure~\ref{fig:IV_pulses}, all devices but one exhibit clear superconducting transitions. From approximate design expectations of saturation powers and comparisons to identical devices measured in a commissioned testbed, we infer a total background load on the order of 10--100~fW. There are clear vibrational and EMI contributions to this background loading, as evidenced through antenna tests and measurements with the pulse tube temporarily powered off; future work will address these.  We also infer parasitic resistances in series with the TES, $R_p$, between 4--7~m$\Omega$ from fits to the region where the TES is superconducting, consistent with expectations from the micro-D connectors and within the acceptable range for $\mathcal{O}(100)$~m$\Omega$ operating resistance sensors.

\par
When biased into their superconducting transition, we see clear correlated background pulses in the two QET channels which share an absorber, as is expected of athermal phonon bursts \cite{anthonypetersen2025}.  An example is shown in the right panel of Figure~\ref{fig:IV_pulses}.  These are likely a mix of radiogenic decays, cosmic rays, and LEE events, though our initial setup does not include a calibration source for further verification.  Currently, the rate of visually resolvable LEE-like events near the noise floor is $\gtrsim$100~Hz, which may be due to the adhesive mounting scheme of the TES chips; future work with improved chip mounting is expected to reduce this greatly.

\par
A significant concern for low-threshold TES sensors is that parasitic leakage of the microwave resonator probe tones could leak into the TES.  
Analytic models of the multiplexer chip's lumped-element components suggest that this leakage can reach the fW scale \cite{mates2011} -- acceptable for $\sim$pW sensors used in mm-wave/x-ray/gamma-ray applications, but problematic for envisioned $\lesssim$fW athermal phonon sensors.  We therefore performed a dedicated measurement to place an upper bound on the parasitic microwave loading in one QET channel. We biased the TES to $R = 0.1R_N$, fixed the RFSoC-generated probe tone, and varied the total RF power using a programmable variable attenuator\footnote{MiniCircuits RCDAT-8000-60} as we recorded a stream of data. The inline attenuation was varied between 0 and 20 dB, over which the measured phase change at the $\approx$4.6 GHz resonator frequencies was $\leq 20^\circ$.  While an ideal $\mu$mux channel with fixed-tone readout is in principle insensitive to phase changes on the readout line, we observe a small baseline shift in practice.  

\begin{figure}[htbp]
\centering
\includegraphics[width=\linewidth]{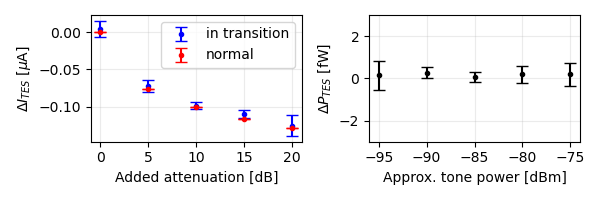}
\caption{Left: measured baseline readout signal shifts due to varying the tone power and phase.  Right: change in power sensed by the TES, inferred through the nominal responsivity.  The uncertainty is presently dominated by the high background pulse rate in our setup, as is evident from the significantly larger uncertainty on the current measurement while in the transition than above it.}
\label{fig:readout_load}
\end{figure}

We calibrated this out using measurements taken with the TES biased above its superconducting transition (where its power-to-current responsitivity $dI/dP$ is zero), as shown in Figure~\ref{fig:readout_load} (left). The right panel shows the inferred change in power incident on the TES from the readout after removing this calibrated phase effect and converting to power assuming the TES responsivity follows the standard DC relation $dI/dP = -1/(I(R-R_p-R_{sh}))$.  We see no evidence of a changing power due to this attenuation sweep, with uncertainties of $\approx$1 fW dominated by the contaminating influence of the high background pulse rate.  Improved precision is anticipated with better chip mounting in the future.

\section{Discussion and Conclusion}
\label{sec:discussion}
\par
Though this first iteration of prototype multiplexed readout for athermal phonon dark matter TES detectors works well in many regards, a number of future improvements are still identifiable at this stage. As discussed in Section~\ref{sec:readout_checks}, the best noise is achieved with a flux ramp rate, and hence sample rate, of $\approx$7 kHz. This is relatively slow for envisioned dark matter experiments that require higher sampling rates to resolve the rising edge of pulses.  

Future work will transition to existing chip designs with 3--10x greater bandwidth \cite{mates2017, Szypryt_2023} to enable higher sample rates.  Additionally, the noise induced by the readout is currently somewhat high: the 34~pA/$\sqrt{\mathrm{Hz}}$ level measured in Section~\ref{sec:readout_checks} would increase the total noise-equivalent power of a modern low-threshold athermal phonon TES dark matter detector by 15\%.\footnote{using the "left" device described in \cite{anthonypetersen2025} as an example, assuming it is biased to 10\% of its normal resistance}  This is an open problem, but as discussed recently in \cite{iachellini2026}, various options including stronger input coupling, tone tracking with overcoupled SQUIDs, and parametric preamplification offer potential paths to reducing the readout noies in the future.  Lastly, though the total RF leakage-induced loading on the TESs is sufficiently small in our current setup, the possibility remains that the parasitic wiring impedances in the $\sim$10 cm twisted pair cable between the multiplexer chip and TES chips are larger than a carefully designed lithographed fanout would have.  The twisted pair may therefore be artificially protecting the TESs by filtering out high-frequency resonator tones in a way that may not mimic a realistic final experiment design.  If this is the case, we expect the inline filtering scheme discussed in Section~\ref{sec:experimental_preparation} to be of use.

\par
In this work, we have presented R\&D progress towards practical and highly scalable multiplexed readout for future athermal phonon TES-based detector arrays.  Specifically, we have established basic compatibility between $\mu$mux readout and $\gtrsim$fW $P_{sat}$ TESs, with a circuit topology that is plausible for future sub-GeV dark matter experiments.  Furthermore, we verified key readout performance metrics for the system as implemented, bounded the level of parasitic readout-induced TES background loading, and identified inline microwave filtering options.  While the proof-of-concept demonstration here uses a single multiplexer chip with resonators spanning a 150 MHz band to read out only 6 TESs, it could be directly scaled to $\gtrsim$1000 TESs per coaxial cable by daisy-chaining multiplexer chips to fill out the full 4--8 GHz octave.  We expect this work, and in particular the treatment of parasitic microwave leakage, to directly inform the design of future highly multiplexed athermal phonon detector arrays.

\section*{Acknowledgments}
This work was supported by the Office of Science, Office of High Energy Physics, of the U.S. Department of Energy under contract no. DE-AC02-05CH11231. OpenAI's ChatGPT and Google Gemini were used for manuscript proofreading and editing, and Anthropic's Claude was used to assist with software execution.

\newpage 
\bibliographystyle{IEEEtran}
\bibliography{IEEEabrv,main}

\end{document}